\documentclass[preprint,noshowpacs,preprintnumbers,amsmath,amssymb]{revtex4}

\usepackage{amsfonts,amsmath,amssymb}
\usepackage{graphicx}
\usepackage{bm} 
\usepackage{mathrsfs}
\usepackage{textcomp}
\usepackage{subfigure}

\newlength{\figwidth}
\newlength{\figwidthsmall}
\begin{document}


\title{Photoelectron angular distributions from strong-field ionization of oriented molecules}

\author{Lotte Holmegaard$^1$}%
\author{Jonas L. Hansen$^2$}%
\author{Line Kalh{\o}j$^1$}%
\author{Sofie Louise Kragh$^1$}%
\author{Henrik Stapelfeldt$^{1,2}$}%
\email[Corresponding author: ]{henriks@chem.au.dk}%
\affiliation{$^1$\,Department of Chemistry, Aarhus University, 8000 Aarhus C, Denmark \\
   $^2$\,Interdisciplinary Nanoscience Center (iNANO),  Aarhus University, 8000 Aarhus C,
   Denmark}

\author{Frank Filsinger}%
\author{Jochen K\"upper}%
\email[Corresponding author: ]{jochen@fhi-berlin.mpg.de}%
\author{Gerard Meijer}%
\affiliation{Fritz-Haber-Institut der Max-Planck-Gesellschaft, Faradayweg 4-6, 14195 Berlin,
   Germany}

\author{Darko Dimitrovski}
\author{Mahmoud Abu-samha}
\author{Christian P.J. Martiny}
\author{Lars Bojer Madsen}
\email[Corresponding author: ]{bojer@phys.au.dk}
\affiliation{Lundbeck Foundation Theoretical Center for Quantum
System Research, Department of Physics and Astronomy, Aarhus University, 8000 Aarhus C, Denmark}

\date{\today}

\maketitle



\textbf{The combination of photoelectron spectroscopy and ultrafast light sources is on track to set new standards for detailed interrogation of dynamics and reactivity of molecules \cite{stolow:2006:science,Bostedt:PRL:2008,Noller:JACS:2008,meckel:Science:2008,bisgaard:Science:2009,Santra:PRL:2009,Akagi:Science:2009}.  A crucial prerequisite for further progress is the ability to not only detect the electron kinetic energy, as done in traditional photoelectron spectroscopy, but also the photoelectron angular distributions (PADs) in the molecular frame \cite{stolow:2006:science,Kumarappan:PRL:2008,meckel:Science:2008,bisgaard:Science:2009,Akagi:Science:2009,Landers:PRL:2001,Golovin:JCP:2005}. Until recently the only method relied on determining the orientation of the molecular frame after ionization \cite{SHIGEMASA:PRL:1995,stolow:2006:science,Rolles:PRL:2008,Zimmermann:NPHYS:2008}. This requires that ionization leads to fragmentation thereby limiting both the species and the specific processes that can be studied. An attractive alternative is to fix the molecular frame prior to ionization. The only demonstrations  hitherto involved aligned small linear unpolar molecules \cite{Kumarappan:PRL:2008,meckel:Science:2008,bisgaard:Science:2009}. A decisive milestone is extension to the general class of polar molecules.
Here carbonylsulfide (OCS) and benzonitrile (C$_7$H$_5$N) molecules, fixed in space by combined laser and electrostatic fields, are ionized with intense, circularly polarized, 30 femtosecond laser pulses. For 1-dimensionally oriented OCS the molecular frame PADs exhibit pronounced anisotropies, perpendicular to the fixed permanent dipole moment, that are absent in PADs from randomly oriented molecules. For 3-dimensionally oriented C$_7$H$_5$N additional striking structures appear due to suppression of electron emission in nodal planes of the fixed electronic orbitals. Our theoretical analysis, relying on tunneling ionization theory~\cite{Perelomov:1966:JETP, Ammosov:1986:JETP},  shows that the PADs reflect nodal planes,  permanent dipole moments and polarizabilities of both the neutral molecule and its cation. The calculated results are exponentially sensitive to changes in these molecular properties thereby pointing to exciting opportunities for time-resolved probing of valence electrons dynamics by intense circularly polarized pulses. Molecular frame PADs from oriented molecules will prove important in other contexts notably in emerging free-electron-laser studies where localized inner shell electrons are knocked off by x-ray pulses.}

Experimentally a target of adiabatically aligned and oriented molecules is created by the combined action of a 10 nanosecond laser pulse and a weak static electric field \cite{friedrich:1999:jcp,Holmegaard:PRL102:023001}. Here alignment refers to confinement of molecule-fixed axes along laboratory fixed axes, and orientation refers to the molecular dipole moment pointing in a particular direction \cite{stapelfeldt:2003:rmp}. Before reaching the interaction point with the laser pulses and the static field the molecules are selected in the lowest-lying rotational quantum states by an electrostatic deflector \cite{Filsinger:JCP:2009}. Hereby alignment and orientation is optimized, which is crucial for observation of the molecular frame PAD effects discussed next. The degree of alignment and orientation is initially measured by Coulomb exploding the molecules using an intense femtosecond (fs) probe laser pulse (Supplementary Information, SI).

\begin{figure}
    \centering
   \includegraphics[width=\figwidth]{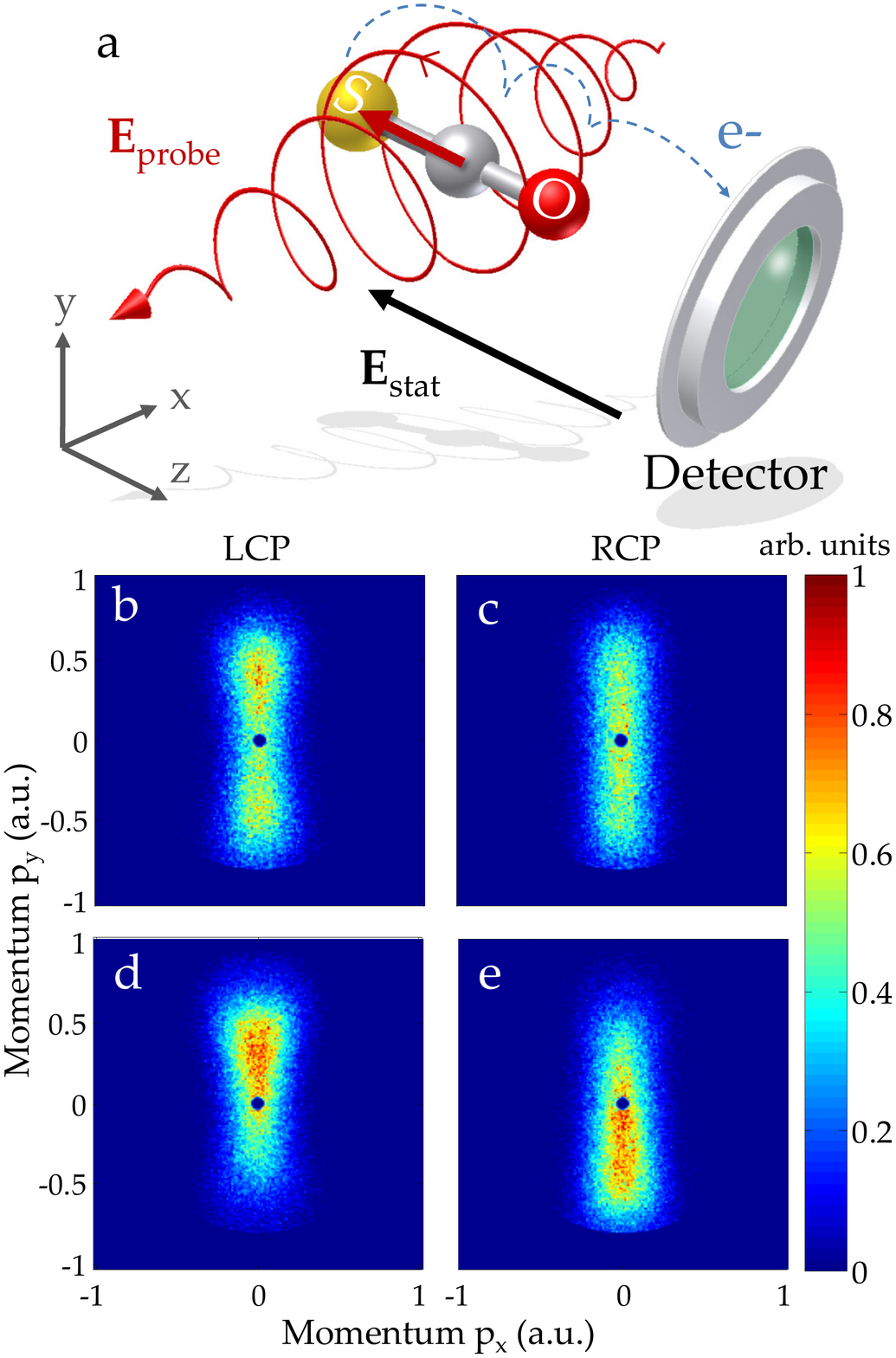}
 \caption{\textbf{Experimental photoelectron images from OCS.}
 \textbf{a}, Schematic of the experimental setup showing an OCS molecule oriented with its permanent dipole moment (bold red arrow) pointing in the direction of the static electric field, ${\mathbf E}_\text{stat}$. The LCP probe pulse ionizes the molecule and imparts an upward momentum to the freed electron resulting in recording on the upper part of the detector (see text for details). \textbf{b}, Two-dimensional momentum image of electrons produced when a (nearly) randomly oriented sample of OCS molecules are ionized by the LCP probe pulse. The polarization plane of the probe pulse is perpendicular to the image (detector). \textbf{c}, Same as \textbf{b} but for a RCP probe pulse. \textbf{d, e}, as \textbf{b} and \textbf{c} but with the OCS molecules oriented as in \textbf{a} by the alignment pulse polarized perpendicular to the image plane.}
  \label{fig1}
\end{figure}

For the PAD experiments a circularly polarized, 30 fs probe pulse, centered at 800 nm, is used. First, the linear OCS molecule is studied. At a peak intensity $\simeq 2.4 \times 10^{14}$ W/cm$^2$, OCS only undergoes single ionization with essentially no fragmentation. The intensity puts the dynamics in the tunneling regime~\cite{Keldysh} (SI) and the circular polarization ensures that no recollision of the freed electron with its parent ion occurs. Both conditions are important for the interpretation and modeling of the observed PADs. The linear polarization of the alignment laser is parallel to the static field axis (Fig. \ref{fig1}a). Hereby, the OCS molecules are strongly confined along this axis with a degree of orientation corresponding to about 80\% of the molecules having their O-end facing the detector and 20\% facing oppositely.

The electron images are shown in Fig. \ref{fig1}. Applying only the probe pulse (Fig. \ref{fig1}b and c) the electrons emerge in a stripe parallel to the polarization plane (y,z) of the probe pulse for both left and right circularly polarized (LCP and RCP) pulses. When the molecules are 1-dimensionally (1D) aligned, and oriented as shown in Fig. \ref{fig1}a, a strong up-down asymmetry is observed (Fig. \ref{fig1}d and e). The asymmetry reverses as the helicity of the probe pulses is flipped. For LCP (RCP) probe pulses the number of electrons detected in the upper part compared to the total number in the image is $\sim$64\%  (39\%). Without the alignment pulse only a very weak asymmetry is seen. It may result from very mild alignment and orientation due to the interaction between the permanent dipole moment and the static field \cite{Nevo:PCCP:2009}.

To explain the experimental findings we model the ionization process by modified tunneling theory. The model (SI) is based on the static tunneling rate \cite{Perelomov:1966:JETP,Ammosov:1986:JETP} for an s-state
with the binding energy of the
highest occupied molecular orbital (HOMO) in OCS, taking saturation \cite{Tong:2005:JPB} into account  and, importantly, including Stark shifts of both OCS and OCS$^+$ energy levels due to the interaction between the probe laser field, ${\mathbf E}_\text{probe}$, and the permanent and induced dipole moments.
The Stark shifts lead to an effective ionization potential, $I_\text{p}^\text{eff}(\theta)$, used in the tunneling model, given by:
\begin{eqnarray}
I_\text{p}^\text{eff}(\theta)&=& I_\text{p0} + (\mu^{\text{OCS$^+$}} - \mu^{\text{OCS}}) E_\text{probe} \cos \theta \\ \nonumber
&+& \frac{1}{2} E_\text{probe}^2   [ \{ ( \alpha_{\parallel}^\text{OCS} -\alpha_{\parallel}^\text{OCS$^+$} ) - ( \alpha_{\perp}^\text{OCS} -\alpha_{\perp}^\text{OCS$^+$} )\} \cos^2\theta + (\alpha_{\perp}^\text{OCS} - \alpha_{\perp}^\text{OCS$^+$})].
\end{eqnarray}
Here $ \mu^{\text{OCS}}$ ($\mu^{\text{OCS$^+$}}$) is the permanent dipole moment of OCS (OCS$^+$), $\alpha_{\parallel}^\text{OCS}$ and $\alpha_{\perp}^\text{OCS}$ ($\alpha_{\parallel}^\text{OCS$^+$}$ and $\alpha_{\perp}^\text{OCS$^+$}$) the polarizability components of OCS (OCS$^+$) parallel and perpendicular to the internuclear axis, $I_\text{p0}$ the ionization potential of OCS in the absence of any external fields, and $\theta$ the polar angle between the instantaneous direction of the  circurlarly polarized probe field and the z-axis (Fig. 1a). Since the ionization (tunneling) rate depends exponentially on the effective ionization potential \cite{Ammosov:1986:JETP}  equation (1) shows directly that the oriented OCS molecules may have an asymmetric ionization probability depending on whether the probe field has a component parallel or anti-parallel to the permanent dipole moment.

\begin{figure}
    \centering
   \includegraphics[width = 160 mm]{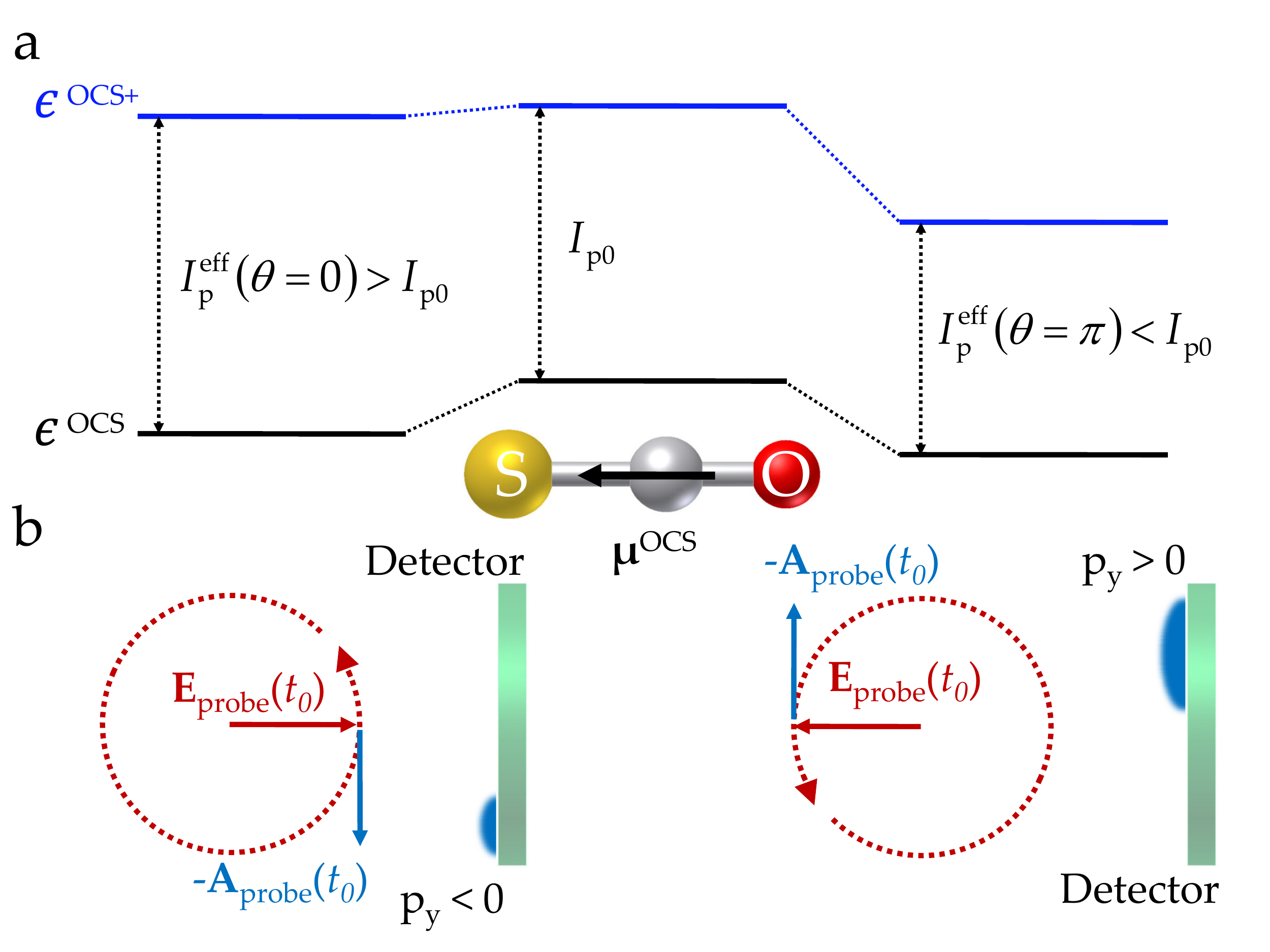}
 \caption{\textbf{Principle of OCS ionization and electron emission.}
 \textbf{a}, Sketch of the effective ionization potential for ${\mathbf E}_\text{probe}$ anti-parallel to ${\bm \mu}^\text{OCS}$ (left), no ${\mathbf E}_\text{probe}$ (middle) and ${\mathbf E}_\text{probe}$ parallel to ${\bm \mu}^\text{OCS}$ (right). The energy levels of OCS (OCS$^+$) are shown by bold black (blue) horizontal lines. \textbf{b}, Illustration of the momentum transfer to the electron due to the force from  the LCP probe field. In the tunneling process the electron escapes oppositely to the instantaneous direction of the probe field and acquires the final momentum, $-{\mathbf A}_\text{probe}(t_0)$. When ${\mathbf E}_\text{probe}$ is parallel (anti-parallel) to ${\bm \mu}^\text{OCS}$ the electron receives an upward (downward) momentum, i.e., p$_y~>~0$ (p$_y~<~0$). The ionization rate is higher in the parallel case causing more upward than downward electrons.}
  \label{fig2}
\end{figure}

\begin{figure}
    \centering
   \includegraphics[width = 160 mm]{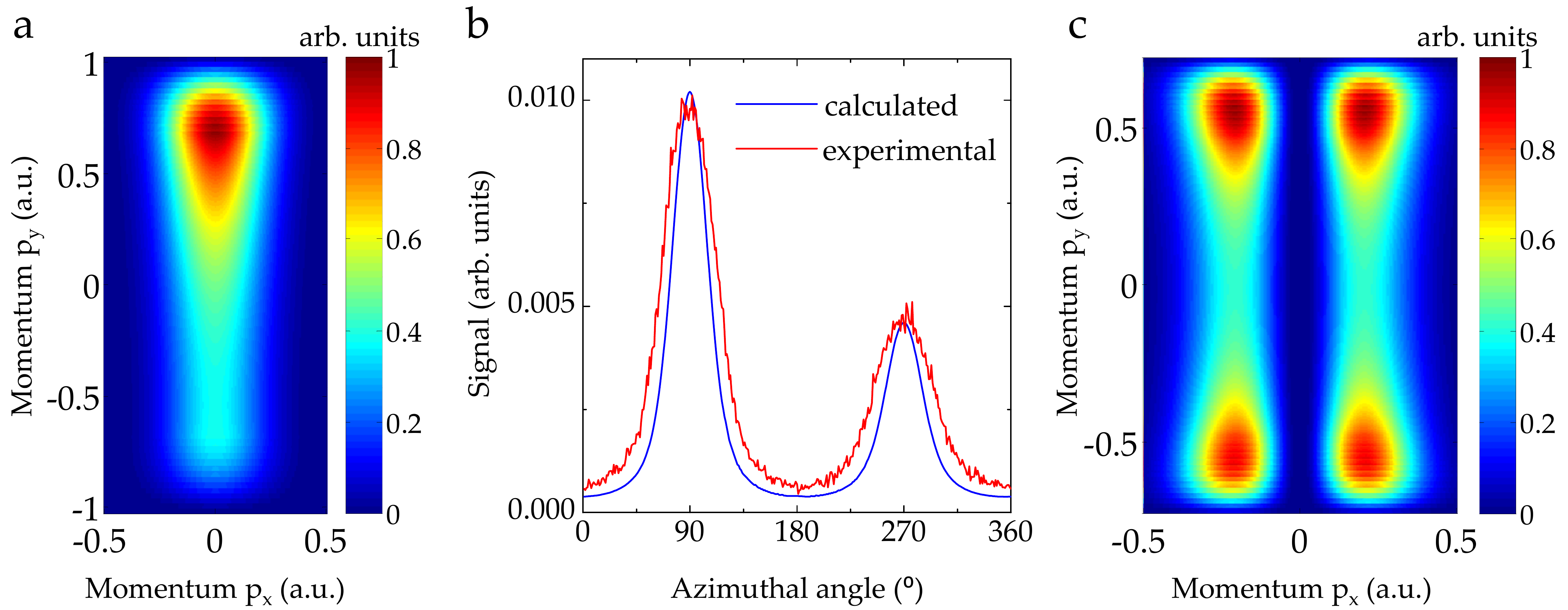}
 \caption{\textbf{Calculated photoelectron images and angular distributions}
 \textbf{a}, Calculated 2-dimensional electron momentum distribution for ionization of 1D aligned and oriented OCS employing a LCP probe pulse similar to the one used in the experiment (Fig. 2d).  \textbf{b}, Calculated (blue curve) angular distribution of the photoelectrons from the distribution in {\bf a}. The red curve is the experimental angular distribution obtained from the image displayed in Fig. 2d. \textbf{c}, Calculated 2-dimensional electron momentum distribution for ionization from the HOMO and HOMO-1 of perfectly 3D aligned and oriented C$_7$H$_5$N employing a LCP probe pulse similar to the one used experimentally (Fig. 4e).}
  \label{fig3}
\end{figure}

The situation is illustrated in Fig. \ref{fig2}a. It shows that $I_\text{p}^\text{eff}(\theta = \pi)$ is smaller than $I_\text{p}^\text{eff}(\theta = 0)$, i.e., the ionization rate is larger for ${\mathbf E}_\text{probe}$ parallel rather than anti-parallel to the permanent dipole moment. More generally equation (1) describes that $I_p$ is smallest, and therefore ionization most probable, for $\pi/2 \leq \theta \leq 3\pi/2$, corresponding to the half part of the optical period where the probe field has a component pointing towards the S-end. This angular dependence of the ionization probability causes a forward-backward asymmetry of the electron emission from the molecule with more electrons ejected when the field points in the direction of the permanent dipole moment. Before the electrons reach the detector they are subject to the force from the remaining part of the strong probe field, which leads to the final momentum distribution: ${\mathbf p}_f = - \vert e \vert \int_{t_0}^\infty {\mathbf E_\text{probe}}(t) dt = - \vert e \vert {\mathbf A}_\text{probe}(t_0)$, where $t_0$ is the instant of ionization. The vector potential, ${\mathbf A}_\text{probe}(t)$, for LCP (RCP) advances the field by a phase of $\pi/2$ (-$\pi/2$) causing the forward-backward asymmetry in the ionization step to be transferred into an up-down asymmetry in the final momentum distribution (along the y-direction, Fig. \ref{fig1}a). This is illustrated in Fig. \ref{fig2}b. The calculated momentum distribution, projected onto the plane corresponding to the detector, is shown in Fig. \ref{fig3}a for LCP probe pulses with the same characteristics as the experimental pulses. (The result with RCP pulses is identical except for having the opposite up-down asymmetry). Focal volume effects are included \cite{Wang:2005:OPL} and an orientation of 80/20 based on the experimental findings is assumed. The similarity with the measurements (Fig. \ref{fig1}d) is clear. In particular, the theoretical up/total ratio of 65 \%, compares very well with the measured value (64 \%). A more quantitative comparison is provided in Fig. \ref{fig3}b, where the experimental and numerical angular distribution for LCP is plotted, obtained by radially integrating the images in Fig. 2 and 3.  The agreement is gratifying.

\begin{figure}[ht]
    \centering
 \includegraphics[width = 160 mm]{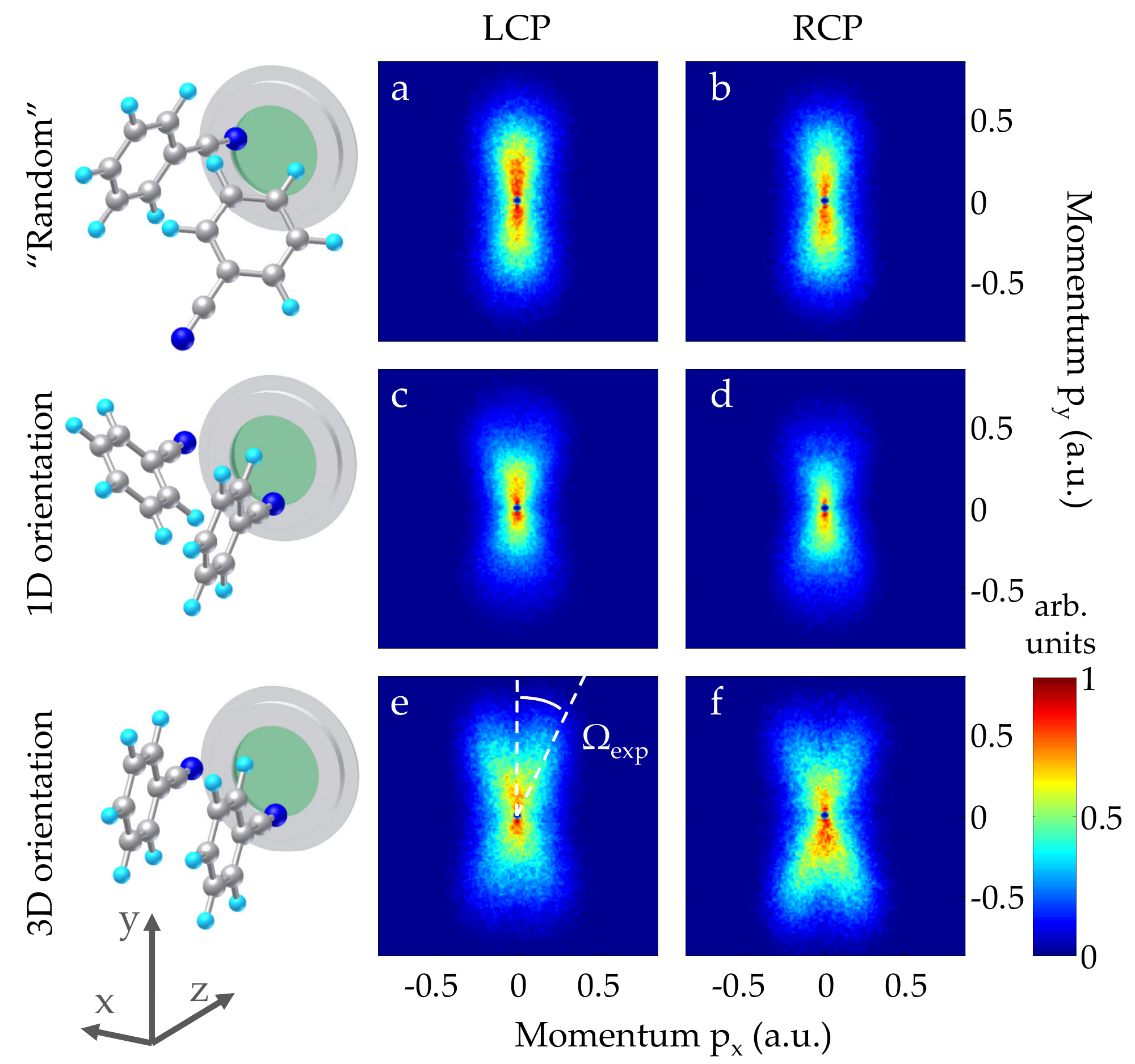}
 \caption{\textbf{Experimental photoelectron images from benzonitrile.}
Two-dimensional momentum image of electrons produced when C$_7$H$_5$N molecules are ionized by a LCP probe pulse \textbf{(a, c, e)} or RCP probe pulse \textbf{(b, d, f)}. In \textbf{a} and \textbf{b} the molecules are essentially randomly oriented (no alignment pulse). In \textbf{c} and \textbf{d} the molecules are 1D oriented (linearly polarized alignment pulse) and in \textbf{e} and \textbf{f} the molecules are 3D oriented (elliptically polarized alignment pulse). The intensity of the probe pulse is $1.2 \times10^{14}$~W/cm$^2$. In {\bf e} the experimental off-the-nodal-plane angle, ${\Omega}_\text{exp} \simeq 18^\circ \pm 1^\circ$ is shown. See  equation (2) for our analytical expression.}
  \label{fig4}
\end{figure}

To illustrate the potential of our method to more complex molecules experiments were conducted on C$_7$H$_5$N. As in OCS, Coulomb explosion imaging measurements confirmed that high degrees of alignment and orientation are achieved (SI). Figure \ref{fig4} displays the photoelectron distributions. Even without the alignment pulse an up-down asymmetry (Fig. \ref{fig4}a and b) is observed. Although weak, it is more pronounced than in the case of OCS (Fig. \ref{fig1}b and c), which we ascribe to stronger 1D orientation, induced by the static field \cite{Nevo:PCCP:2009}, because C$_7$H$_5$N has a permanent dipole moment that is 6.3 times larger than that of OCS and the static field used is 35 \% larger. When the 1D alignment pulse, linearly polarized along the static field axis, is included the asymmetry increases (Fig. \ref{fig4}c and d) due to stronger alignment and orientation. For LCP the up/total ratio is $\sim$55\%, for RCP it is $\sim$44\%. Unlike OCS,  C$_7$H$_5$N is an asymmetric top molecule and the linearly polarized alignment pulse only confines the C-CN symmetry axis of the molecule, while the benzene ring remains free to rotate about this axis. Additional confinement of the molecular plane, and thus 3D alignment and orientation, is obtained by using an elliptically polarized alignment pulse \cite{larsen:2000:prl,Nevo:PCCP:2009} with the major axis along the static field axis and the minor axis vertical in the images. The total intensity, $7\times10^{11}$~W/cm$^2$, is the same as for the linearly polarized pulse and the intensity ratio between the major and minor axis is 3:1. The resulting electron distributions (Fig. \ref{fig4}e and f) have essentially the same up-down asymmetry as in panel c and d (since the degree of 1D alignment and orientation is essentially unchanged), but  they exhibit striking new structures that are not seen in the PADs from 1D oriented molecules. In particular, electron emission in the polarization plane, coinciding with the nodal plane of the HOMO (and the HOMO-1), is suppressed.  Such effects were predicted previously but never observed~\cite{kjeldsen:2005:pra}.

To explain the experimental findings for 3D oriented C$_7$H$_5$N our theoretical model is extended. By contrast to OCS the HOMO of C$_7$H$_5$N has a nodal plane (the molecular plane) and therefore, we model the initial state of a molecule, perfectly 3D oriented in the way shown in the lower panel of Fig. \ref{fig4}, by a simple $2p_x$ orbital with the angular node in the polarization (y,z) plane and the p-lobes parallel to the x-axis. The presence of the nodal plane prevents electron emission in the polarization plane. This is observed in the calculated electron momentum distribution displayed in Fig. \ref{fig3}c for LCP. Initial electron emission in the perpendicular direction is, however, possible and is incorporated by including a factor describing the  momentum distribution in the x-direction in the initial orbital (SI).
The slight asymmetry of the up/total ratio (52 \%) visible in Fig. \ref{fig3}c is still governed by laser-induced Stark shifts, as described for OCS, whereas the observed left-right splitting (along the x-axis) of the momentum distribution projected on the detector plane is due to the presence of the nodal plane. We obtain a new closed analytical expression for the off-the-nodal-plane angle, $\Omega_\text{theo}$ (Fig. 4e), and at the intensity $1.23 \times 10^{14}$ W/cm$^2$,
 \begin{equation}
 \Omega_\text{theo}= \arctan (2 \omega/\sqrt{\pi F_0 \kappa}) \simeq 18.8^\circ,
\end{equation}
 with $E_0$ the peak electric field, $\kappa = \sqrt{2 I_{p0}}$
 in good agreement with the experimental value,
 $\Omega_\text{exp} = 18^\circ \pm 1^\circ$ (Fig. \ref{fig4}e). The experimental observation of a nonzero electron signal in the polarization plane is partly due to nonperfect 3D alignment of the molecule and possibly also due to a contribution from the HOMO-2 whose nodal plane is perpendicular to the laser polarization plane.

Our results show that molecular frame PADs can be obtained for a broad range of species independent of whether 1D or 3D orientation is needed to fully fix a molecule in space. A particularly interesting extension is time-dependent phenomena where a pump pulse initiates a molecular transformation or reaction.  Strong-field ionization
by circularly polarized fs pulses, used here, is sensitive to the charge distribution of the valence electrons and could provide an efficient and ultrafast probe of, for instance, charge migration processes in molecules \cite{Kuleff:CP:2007}. More generally, molecular frame PADs using laser oriented molecules will also be highly relevant for x-ray probing of molecular dynamics where detection of high energy electrons can provide a direct structural diagnostics of the changing molecular species \cite{Krasniqi:PRA:2009}.

\section*{Acknowledgements}
The work was supported by the Danish National Research Foundation, the Lundbeck Foundation and the Carlsberg Foundation. H.S. thanks P. Corkum for discussions.

\section*{Author contributions}
L.H., J.L.H, L.K., S.L.K., H.S., F. F., J.K., and G.M. designed experiments.
L.H., J.L.H, L.K., S.L.K, H.S. performed experiments.
D.D., M. A.-S., C.P.J.M., and L.B.M. developed the theory. All authors were involved in the completion of the manuscript.

\bibliographystyle{nature}

\clearpage

\end{document}